# Magnetism in Mn-Nanowires and –Clusters as δ-doped Layers in Group IV Semiconductors (Si, Ge)

K. R. Simov[1], P. -A. Glans[2], C. A. Jenkins[2], M. Liberati[2], P. Reinke[1]

(1) Department of Materials Science and Engineering, University of Virginia, 395 McCormick Road, Charlottesville, VA 22901, U.S.A.

(2) Advanced Light Source, Lawrence Berkeley National Laboratory, Berkeley, California 94720, U.S.A.

**Abstract**

Mn doping of group-IV semiconductors (Si/Ge) is achieved by embedding a thin Mn-film as a δ-doped layer in group-IV matrix. The Mn-layer consists of a dense layer of monoatomic Mn-wires, which are oriented perpendicular to the Si(001)-(2x1) dimer rows, or Mn-clusters. The nanostructures are covered with an amorphous Si or Ge capping layer, which conserves the identity of the δ-doped layer. The analysis of the bonding environment with STM is combined with the element-specific detection of the magnetic signature with X-ray magnetic circular dichroism. The largest moment (2.5 $\mu_B$/Mn) is measured for Mn-wires, which have ionic bonding character, with an a-Ge overlayer cap; a-Si capping leads to a slightly reduced moment which has its origin in subtle variation of bonding geometry.[1] Our results directly confirm theoretical predictions on magnetism for Mn-adatoms on Si(001).[2] The moment is quenched to 0.5 $\mu_B$/Mn for δ-doped layers, which are dominated by clusters, and thus develop an antiferromagnetic component from Mn-Mn bonding.

**Manuscript**

Spintronics uses the electron spin for device operation, [3-6] and its success hinges on development of materials offering the requisite control of magnetism and spin current. One of the central challenges in the study and use of dilute magnetic semiconductors (SC) is merging spin-controlled and charge-based devices. This can be achieved with magnetically doped SCs [4, 7] where the magnetic response can be controlled via an electric field. Mn is a prime candidate for magnetic doping and has been used successfully in GaAs(Mn), [8, 9] and was proposed as a magnetic dopant in the group-IV SCs Si and Ge. [1, 2, 5, 10-13] However, the Mn-Si and Mn-Ge systems are inherently complex owing to the coexistence of interstitial, and substitutional sites, clusters and compounds.

The wide range of magnetic properties reported in the literature for Mn-doped group IV SCs suggest that we lack control and understanding of critical material aspects and their relation to the magnetic signatures. Recent work on Mn-ion implantation in Si illustrates the complex interplay between different coordination sites, and defect chemistries which have to be considered in the investigation of magnetically doped semiconductors. [14] It is, however, difficult to establish site distributions in bulk materials. Alternatively we use 2D δ-doped layers where targeted synthesis affords good control of dopant positioning. In addition, the formation of silicides and germanides, and a wide range of metastable materials residing in the Ge (or Si) rich region of the phase diagram [12, 15, 16] challenge control of the material structure and to fully understand the connection between bonding environment and magnetism.

Our work combines synthesis of well-defined nanostructures of Mn on Si(001), and element specific detection of the magnetic signature of Mn with XMCD (x-ray circular magnetic dichroism). Mn is deposited on Si(100)-(2x1) surfaces, which serve as a template for monoatomic Mn-wires, which have been studied previously, and Mn-clusters. [17-21] These Mn-nanostructures are characterized with STM (scanning tunneling microscopy) and subsequently capped with an amorphous Si or Ge layer. The capping process preserves the respective Mn-nanostructures and creates a δ-doped Mn-layer as confirmed with XAS (x-ray absorption spectroscopy) and described in the literature. [22] This approach enables us to modulate the Mn-atom nearest neighbor environment and directly correlate with the magnetic signature. Our study connects the Mn-Si bonding with the Si(001)-(2x1) surface and the amorphous cap to the magnetic signature of δ-doped structures.

Monoatomic Mn-wires self-assemble on the Si(001)-(2x1) reconstruction and are shown in Figure 1.[17, 18] The details of the bonding between nanowires and Si surface are best described by a recent model, [21] where the Mn-atoms are integrated in the Si-dimer leading to Mn-wires running perpendicular to the Si-dimer rows with an interatomic Mn-Mn distance of 0.85 nm. According to the Bethe-Slater curve this interatomic Mn-Mn distance should yield ferromagnetic coupling although this zero order approach neglect the Mn-Si interaction. Ultrasmall Mn-clusters are obtained if the defect concentration on the Si(001) surface is above 5%. [18] Figure 1 shows STM images for most of the nanostructures used in the present study. They include high quality wires in Figure 1(a) and (c), mixed nanostructures with a sizable contribution from clusters in Figure 1(b) and (d), and Mn layers with a thickness of 2 ML, where clusters dominates. All STM images were recorded at room temperature using an etched W-tip with a bias voltage of 1.6 V and a feedback current of 0.05 nA. Imaging conditions are given on previous publications. [17, 18]

Mn-deposition and the growth of a Si or Ge capping layer were performed in an ultrahigh-vacuum preparation chamber connected to an Omicron VT-STM (variable temperature scanning tunneling microscope). The Si surface was cleaned by annealing and repeated flashing cycles. [23] Mn and Si are evaporated using an e-beam evaporator (Mantis) with a rate of about $6.5 \times 10^{-3}$ ML/s. One ML Mn corresponds to $1.15 \times 10^{15}$ atoms/cm$^2$, and rates are calibrated with a quartz crystal monitor. Ge is evaporated from a VEECO effusion cell at about 0.04 ML/s. All deposition are performed at room temperature giving amorphous capping layers with a thickness of 10 ML. At intermittent steps in the early stages of capping layer growth, STM measurements were performed to verify that the Mn nanostructures are preserved. The Mn-wires and clusters can be distinguished from the Ge or Si deposit due to a unique contrast variation as a function of bias voltage.[24] The individual samples are labeled as following: the subscript gives the respective layer thickness in ML (monolayer) and the second element indicates the capping layer material e.g. Mn$_{[0.43]}$Si$_{[10]}$ . Angle-resolved XAS measurements were performed after transport of the samples to the Advanced Light Source and show that the Mn-signal attenuation follows the Lambert-Beer law which confirms the stability of the δ-doped Mn-layer in agreement with Ref. 22. The Mn-coverage can therefore be used to calculate the magnetic moments.[25]

XAS, XMCD, and x-ray linear magnetic dichroism (XLMD) were used to study the electronic and magnetic properties of the embedded Mn nanostructures. [26] The measurements were carried out at beamline 6.3.1 of the Advanced Light Source, and only $Mn_{[0.6]}Ge_{[10]}$ was measured at beamline 4.0.2. The $L_{3,2}$ Mn peaks were recorded in the photon energy range from 630-660 eV. A saturation **B**-field of 2 T was applied parallel to the incident x-ray beam and XAS spectra were collected for +/- polarity of the field with respect to the beam direction in the total electron yield mode. The dichroism signal is obtained by switching the **B**-field direction, and the magnetic alignment is confirmed by inverting the x-ray polarization,. A gold mesh with +250 V bias voltage was used to remove contributions from spurious secondary electrons. The small Mn inventory in our samples and the presence of the capping layer required exceptionally long data acquisition times (> 12 hours per sample) to achieve a reasonable signal-to-noise ratio. Due to the very long acquisition times the sample temperatures vary between 20-50 K between samples but were constant within each measurement. The dichroism signal is extracted from XAS spectra measured with opposing magnetic saturation fields after matching the intensity at the $L_3$ pre-edge position, and then analyzed using the sum rules established by Chen et al. [27] The spectra were corrected for measurement geometry, and the number of holes is adjusted at 4.52 following band structure calculations. [28] XMLD spectra were measured with linearly polarized light in the same geometry. VSM (vibrating sample magnetometry) measurements were performed but the signal-to-noise ratio is poor due to the small Mn inventory and the large contribution from the diamagnetic Si-wafer. The magnetic moments derived from the hysteresis loops agreed within 30% of the XMCD data, the lowest Mn concentrations did not yield VSM data. The Curie temperature ($T_{Curie}$) is between 70 and 100 K.

The XAS spectra summarized in Figure 2 have an $L_3$ peak (638 eV) with two shoulders on the high energy side (639.5 and 641 eV), and an $L_2$ peak doublet at 648 and 650 eV. Peak shapes and positions show little variations across all samples and are in agreement with a predominantly $Mn^{2+}$ ($d^5$) bonding state.[1, 29, 30] The difference between tetrahedrally and octahedrally coordinated $Mn^{2+}$ is small [31] and our resolution did not allow to distinguish between them. The peak shape and positions, and specifically the shape of the $L_3$ pre-edge region, are clearly distinct from Mn-oxides with $d^3$ to $d^5$ configurations. Mn thin films and Mn-doped a-Si (Mn>0.5 at%), on the other hand, exhibit significant spectral broadening of the $L_3$ and $L_2$ peaks, which is due to electron delocalization in the Mn-metal, and a 3d-impurity band for the Mn-doped material, respectively.[1] Even the thickest δ-doped Mn-layer with 2 ML equivalent coverage, which consists of "pancake shaped" clusters, does not display metallic behavior. The ionic signature from the hybridization of Mn-3d band with Si-substrate and capping layer dominates the spectral signature. In a weighted superposition of metal and ionic XAS spectra ($Mn^{2+} d^5$, a-$Mn_xGe_{1-x}$ for x=0.05)[1] the overall spectral signature will only visibly change if the metallic contribution exceeds 30% which is an agreement with our interpretation.

The magnetic moments for all samples are shown in Figure 3 and summarized in Table 1, which also includes the orbital and spin contributions. In the majority of samples $M_{spin}$ dominates and the orbital contributions tends to be small. The highest $M_{total}$ is observed for the best nanowire structures: $Mn_{[0.43]}Si_{[10]}$, $Mn_{[0.39]}Ge_{[10]}$ and $Mn_{[0.71]}Si_{[10]}$ while clustering significantly lowers $M_{total}$. In addition, $M_{total}$ given in moment/atom is identical for $Mn_{[0.43]}Si_{[10]}$, and $Mn_{[0.71]}Si_{[10]}$ which both have a predominantly Mn-nanowires (Figures 1(a) and (b)). The magnetic signature of the nanowires falls in line with the theoretical predictions for Mn-adatoms on Si(001) surfaces, which yield between 2.0 and 4.1 $\mu_B$/Mn for different Mn-bonding geometries. The configuration which is closest to the experimental Mn-wire structure (bonding site "H" as defined in Ref. 2 ) has a moment of 3.2 $\mu_B$/Mn. A hybridization between the Mn-nanowire atoms (localized 3d-state) with the Si-p and s-bands stabilizes the ferromagnetic (FM) coupling. The reduced moment for mixed and Mn-cluster rich layers can be attributed to increasing contributions of direct Mn-Mn bonding, which adds an antiferromagnetic (AFM) component leading to overall reduction in $M_{total}$. In addition, $M_{total}$ has already decreased significantly at 50 K indicating a relatively low $T_{Curie}$. XMLD spectra, which reflect antiferromagnetic coupling, support this interpretation.[32] The intensity of $L_3$ peak in the XMLD spectra correlates with the contributions from Mn-clusters relative to Mn-nanowires: the highest intensity in $L_3$ of the XMLD spectra is seen for $Mn_{[2.0]}Ge_{[10]}$ where the XMCD signal already is relatively small, and the XMLD signal disappears for $Mn_{[0.43]}Si_{[10]}$, $Mn_{[0.39]}Ge_{[10]}$ Despite their relatively poor signal-to-noise ratio XMLD spectra support an increasing AFM component for samples rich in Mn-clusters.

The choice of capping layer, a-Si and a-Ge, has also a sizeable impact on the magnetic signature of the δ-doped Mn-layers. $M_{total}$ is larger for samples with an a-Ge cap despite the slightly higher measurement temperature for Mn$_{[0.39]}$Ge$_{[10]}$ compared to Mn$_{[0.41]}$Si$_{[10]}$. The latter is the only sample with a sizeable $M_{orb}$, whose origin is currently not understood. The position of the Mn-atoms is fixed by the Si(001) substrate but the number of nearest neighbors on the side of the cap will be lower for Ge than for Si. The correlation between the moment at the Mn atom and the number of nearest neighbors has been studied for amorphous Si-Mn and Ge-Mn materials, [1] and can be traced back to the differences in the density of states of the two spin channels, which is in turn controlled by the hybridization between SC and Mn states. A larger number of nearest neighbors will lead to a partial quenching of the local Mn moments, which is commensurate with our observation of a lower moment for the a-Si-cap. The complete quenching of the Mn moments and transition to a metallic signature in XAS, which has been observed for a-Si:Mn, is prevented for the nanowires by bonding to the Si(001) surface.

In low-dimensional systems the spin-orbital coupling defines the orientation of the easy magnetic axis. [33, 34] In ultrathin transition metal thin films such as Co embedded in Au, the easy axis can be oriented out of plane due to the spin orbit coupling and associated magnetocrystalline anisotropy. This overrides the shape anisotropy which favors alignment of the moment within the plane of the magnetic layer. Measuring $M_{orb}$ for two angles in Mn$_{[2.0]}$Ge$_{[10.0]}$ yields 0.09±0.02 $\mu_B$/Mn for the normal direction, and 0.01±0.02 $\mu_B$/Mn for 60º with respect to the normal, which is our customary measurement geometry. This is commensurate with a larger spin-orbit coupling and easy axis in the out-of-plane direction. Consequently, the easy magnetic axis is in-plane for the Mn$_{[2.0]}$Ge$_{[10.0]}$ clustered sample. The majority of our samples shows a very small $M_{orb}$ for 60º and thus might also prefer the out-of-plane easy axis. However, this analysis should be taken with a grain of salt do to the overall small values of the orbital moment and limited number of data points.

The combination of atomic level control in the synthesis of magnetic nanostructures and element-specific measurements of magnetic properties yields a new level of understanding in the magnetic doping of group IV semiconductors. In conclusion, we demonstrated the ferromagnetic coupling between Mn-atoms organized within monoatomic chains on the Si(001) surface, although the Curie temperatures remain relatively low. The antiferromagnetic Mn-Mn interaction, which is present in Mn-nanoclusters, depresses $M_{total}$ and confirms that any clustering will diminish magnetic performance of a doped semiconductor. It is also quite remarkable that the impact of the capping layer material is evident in our data, and shows that careful selection of capping layer material can enhance performance. Overall this work, illustrates clearly that the unusual combination of STM and XMCD performed for the same sample is a powerful method for interpreting and understanding magnetism in nanostructures.

**Acknowledgements:** The authors gratefully acknowledge the support of this work by NSF with the award number DMR-0907234, Division of Materials Research (Electronic and Photonic Materials). This work was only possible in collaboration with the scientists and excellent facilities at the Advanced Light Source. The Advanced Light Source is supported by the Director, Office of Science, Office of Basic Energy Sciences, of the U.S. Department of Energy under Contract No. DE-AC02-05CH11231

| Composition | Comment<br>Mn nanostructure T[K], angle to surface normal | $M_{orb}$ ($\mu_B$/Mn) | $M_{spin}$ ($\mu_B$/Mn) | $M_{total}$ ($\mu_B$/Mn) |
|---|---|---|---|---|
| **(a) Mn$_{[0.43]}$Si$_{[10.0]}$** | wires<br>T=20 K, 60° | 0.20 ±0.04 | 1.62 ±0.11 | 1.82 ±0.18 |
| **(b) Mn$_{[0.39]}$Ge$_{[10.0]}$** | wires<br>T=30 K, 60° | 0.02 ±0.02 | 2.48 ±0.21 | 2.50 ±0.24 |
| **(c) Mn$_{[0.71]}$Si$_{[10.0]}$** | mostly wires<br>T=35 K, 60° | 0.06 ±0.01 | 1.95 ±0.31 | 2.01 ±0.35 |
| **(d) Mn$_{[0.6]}$Ge$_{[10.0]}$** | mixed | only XAS available | | |
| **(e1) Mn$_{[2.0]}$Ge$_{[10.0]}$** | clusters<br>T=30 K, 60° | 0.01 ±0.01 | 1.15 ±0.12 | 1.16 ±0.14 |
| **(e2) Mn$_{[2.0]}$Ge$_{[10.0]}$** | clusters<br>T=30 K, 0° | 0.09 ±0.02 | 0.81 ±0.07 | 0.90 ±0.09 |

**Table 1:** Summary of magnetic moments calculated from XMCD spectra using the sum rules. The sample compositions are given in the first column, and nanostructure type, temperature, and angle to the surface normal are summarized in the "Comment" column. $M_{orb}$ is the magnetic orbital moment, $M_{spin}$ is the spin moment, and their sum, the total magnetic moment $M_{total}$ is shown in Figure 3(b). Mn$_{[2.0]}$Ge$_{[10.0]}$ is the only sample where XMCD spectra could be obtained for two different angles given with respect to the substrate normal.

**Figure Captions**

**Figure 1:** STM images of the Mn-nanostructures. These are the samples which were used after capping in the XMCD analysis. Mn can self-assembles on the Si(001)-(2x1) surface into wires (a) and (c), mostly wires with few clusters (b) and (d), and only clusters for coverages exceeding 1 ML (e). Details of the wire formation, atomic scale structure and interplay between wire and cluster formation is given in. [17, 18] The top row of samples (a-b) were capped with an amorphous Si layer with a thickness of 10 ML, and bottom row samples (c-d-e) were capped with an amorphous Ge layer.

**Figure 2:** (a) XAS spectra of the $L_{3,2}$ Mn edge for all samples shown in Figure 1. Sample $Mn_{[0.6]}Ge_{[10.0]}$ was measured at beamline 4.0.2 with a higher energy resolution leading to narrower features in the spectrum. The $L_{3,2}$ intensity as a function of take-off angle for $Mn_{[0.71]}Si_{[10.0]}$ and $Mn_{[2.0]}Ge_{[10.0]}$ follows the Lambert-Beer law if a buried Mn layer is assumed with no Mn contribution within the capping layer or at the surface.

**Figure 3:** XMCD spectra for the $L_3$ Mn peak for the samples shown in Figure 1. The moment $M_{total}$ for all samples is summarized in the graph on the right hand side. The values for the spin and orbital moments are summarized in detail in Table 1. The small Mn inventory combined with attenuation of the signal due to the capping layer contribute to the relatively high noise in the dichroism spectra, which is taken into account in the error bar included in $M_{total}$.

Figure 1

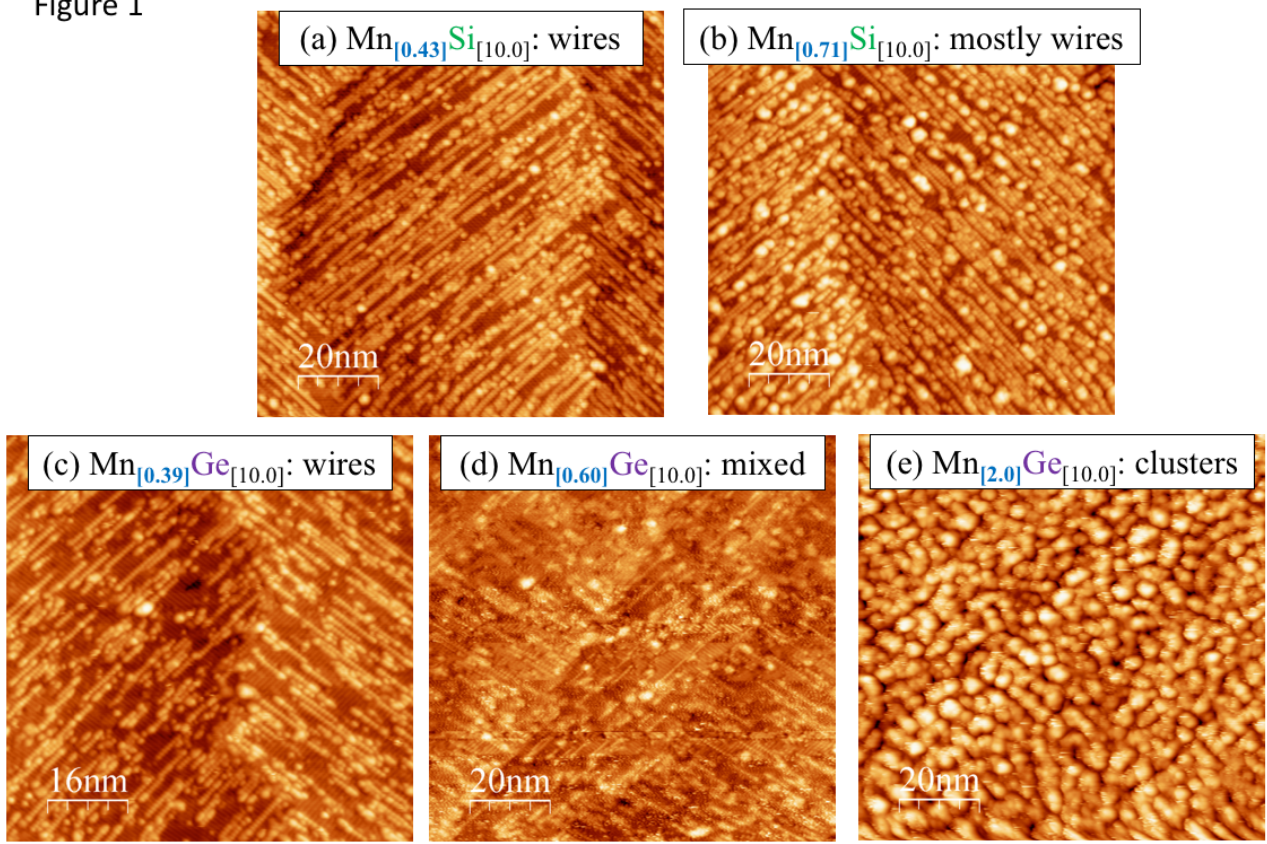

Figure 2

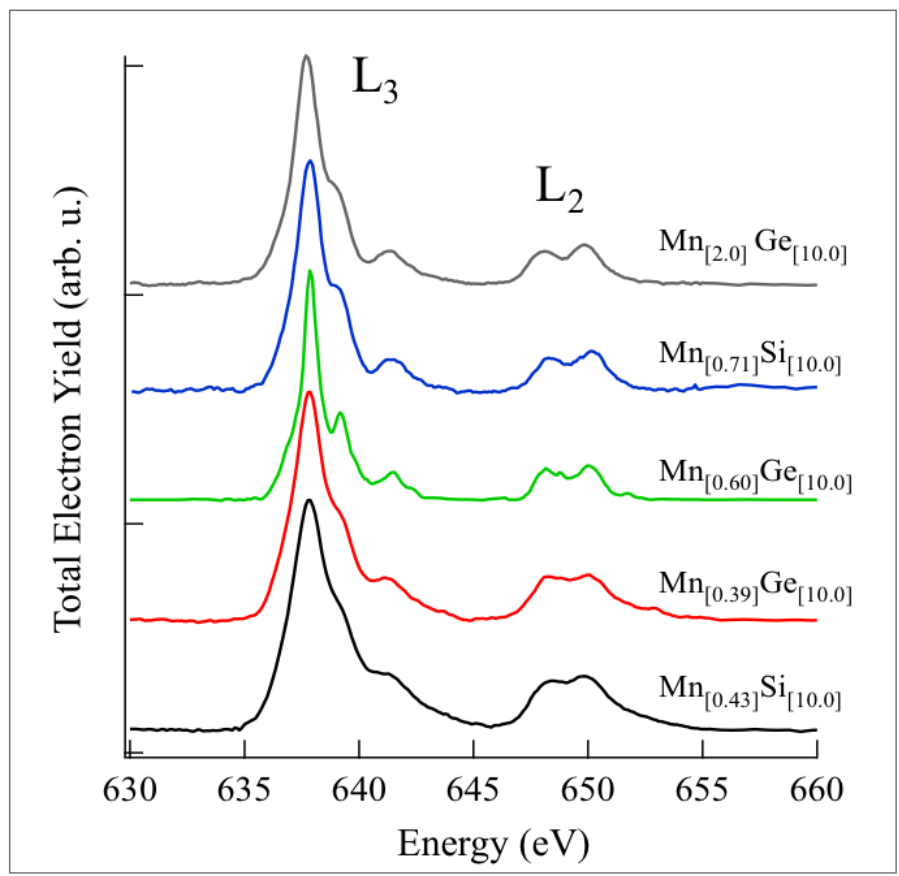

Figure 3

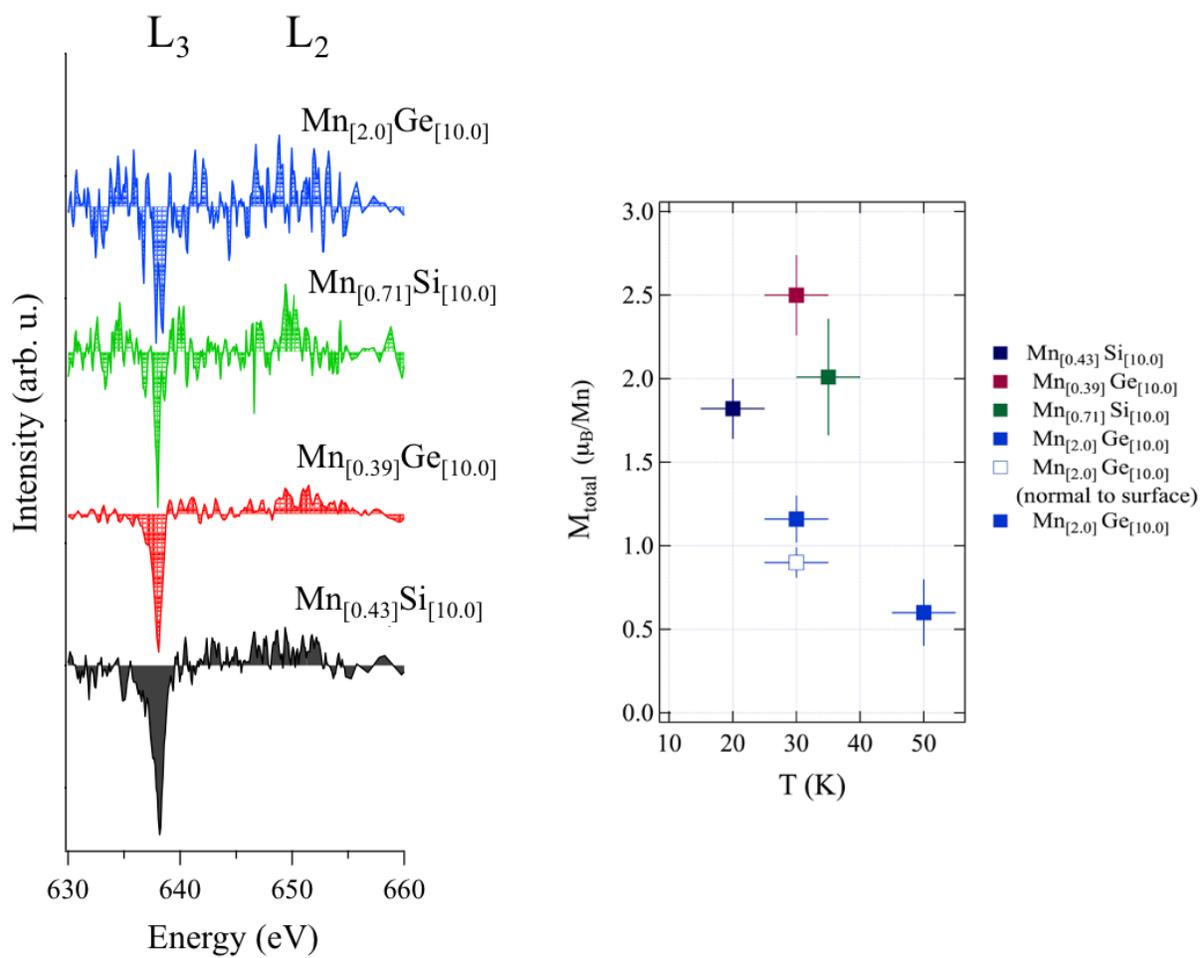

Figure 1

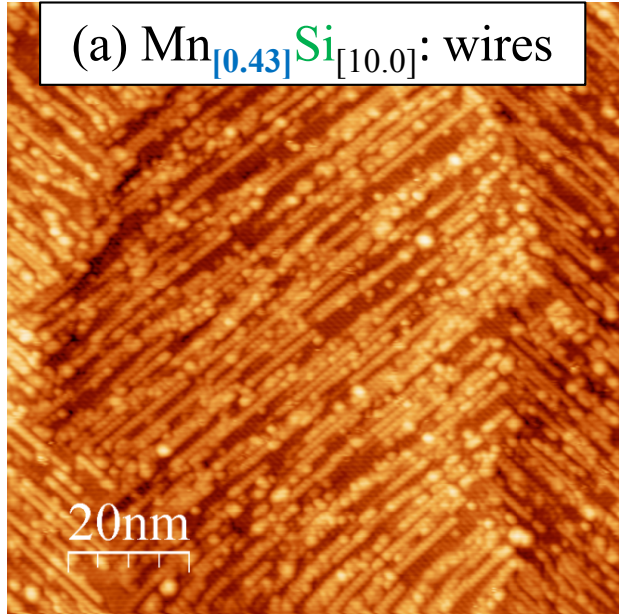
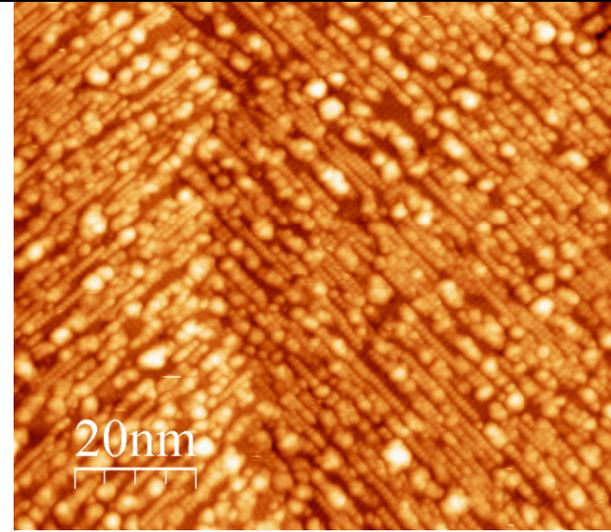
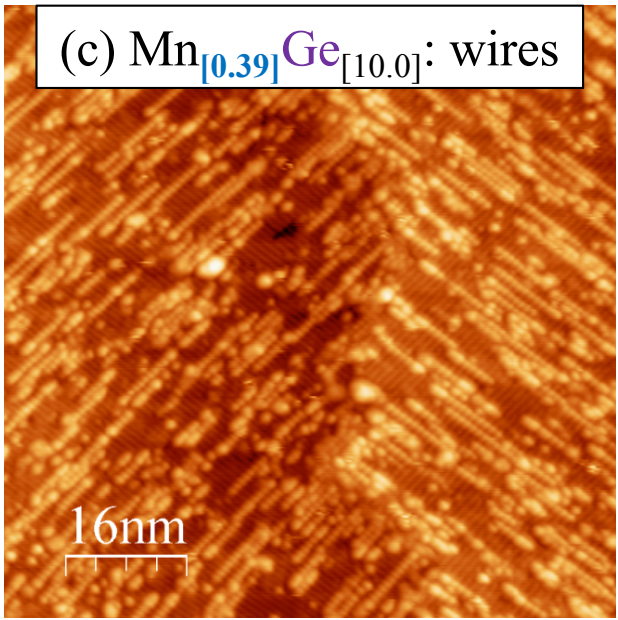
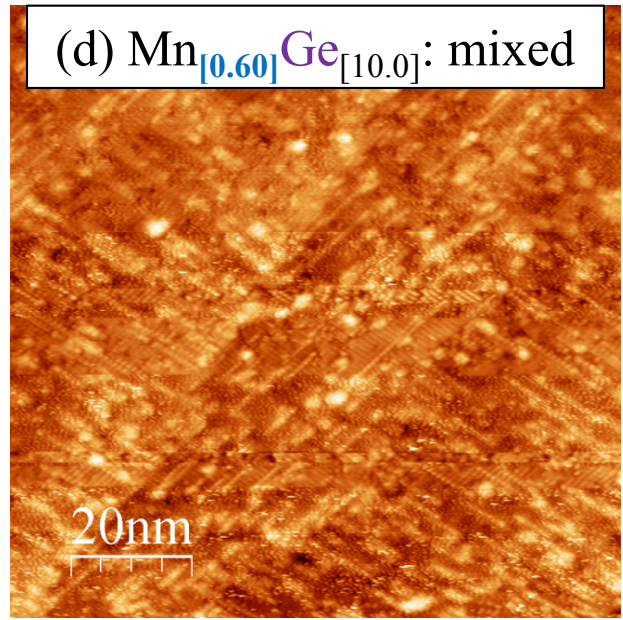
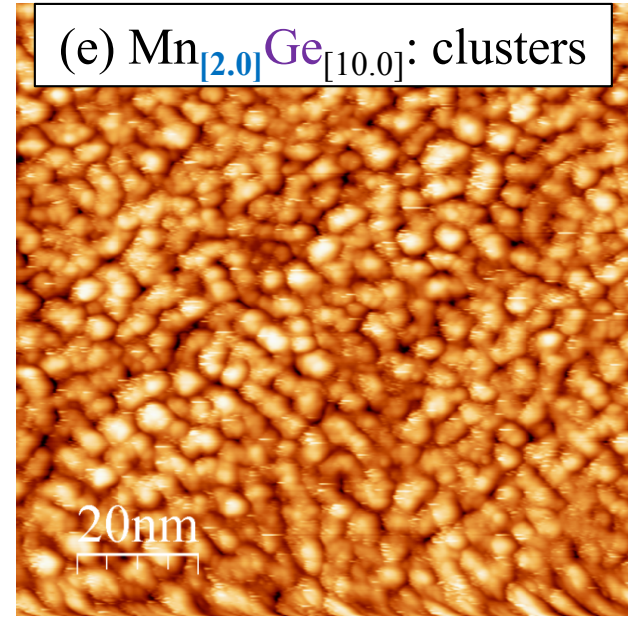

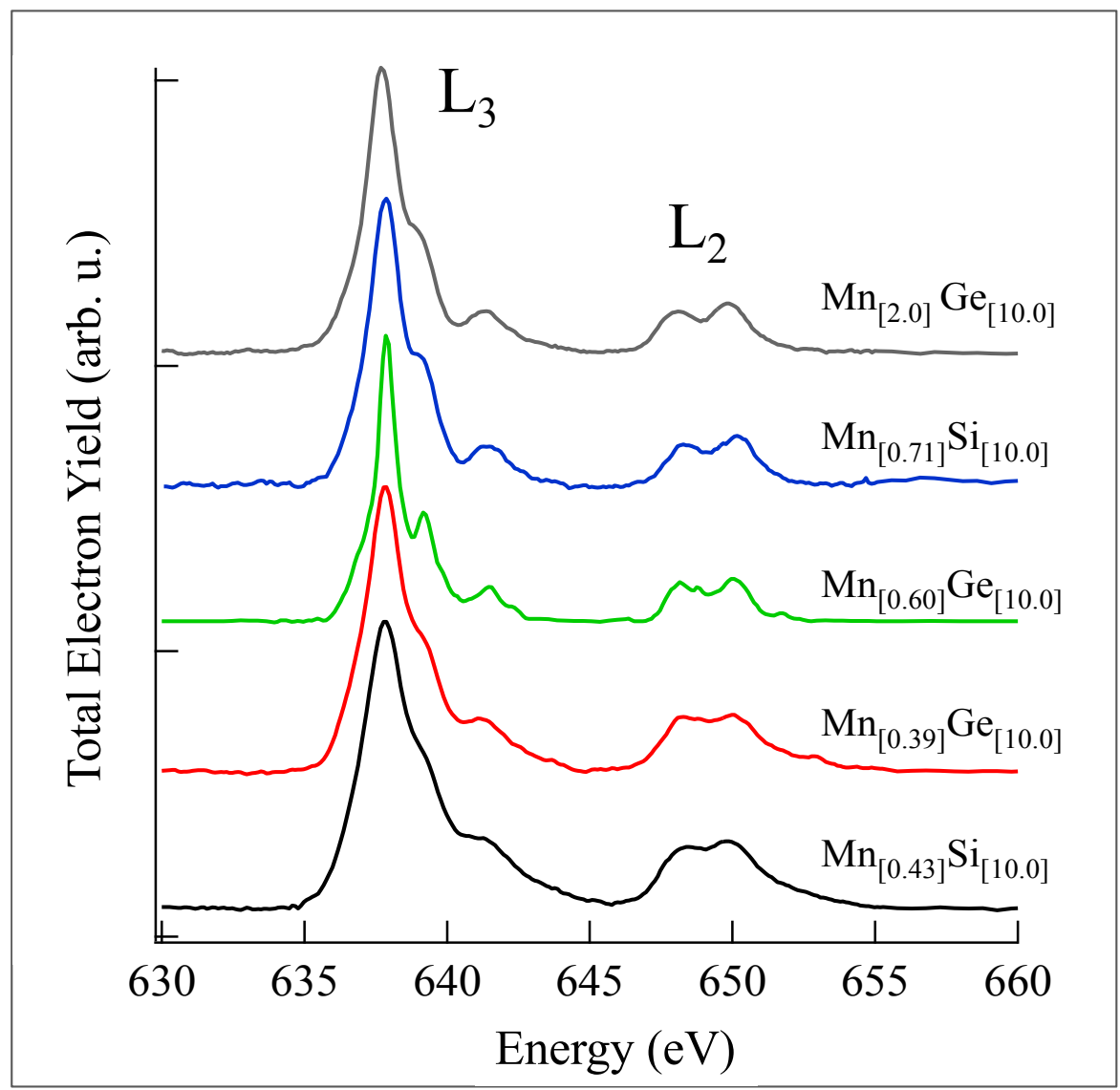

Figure 2

Figure 3

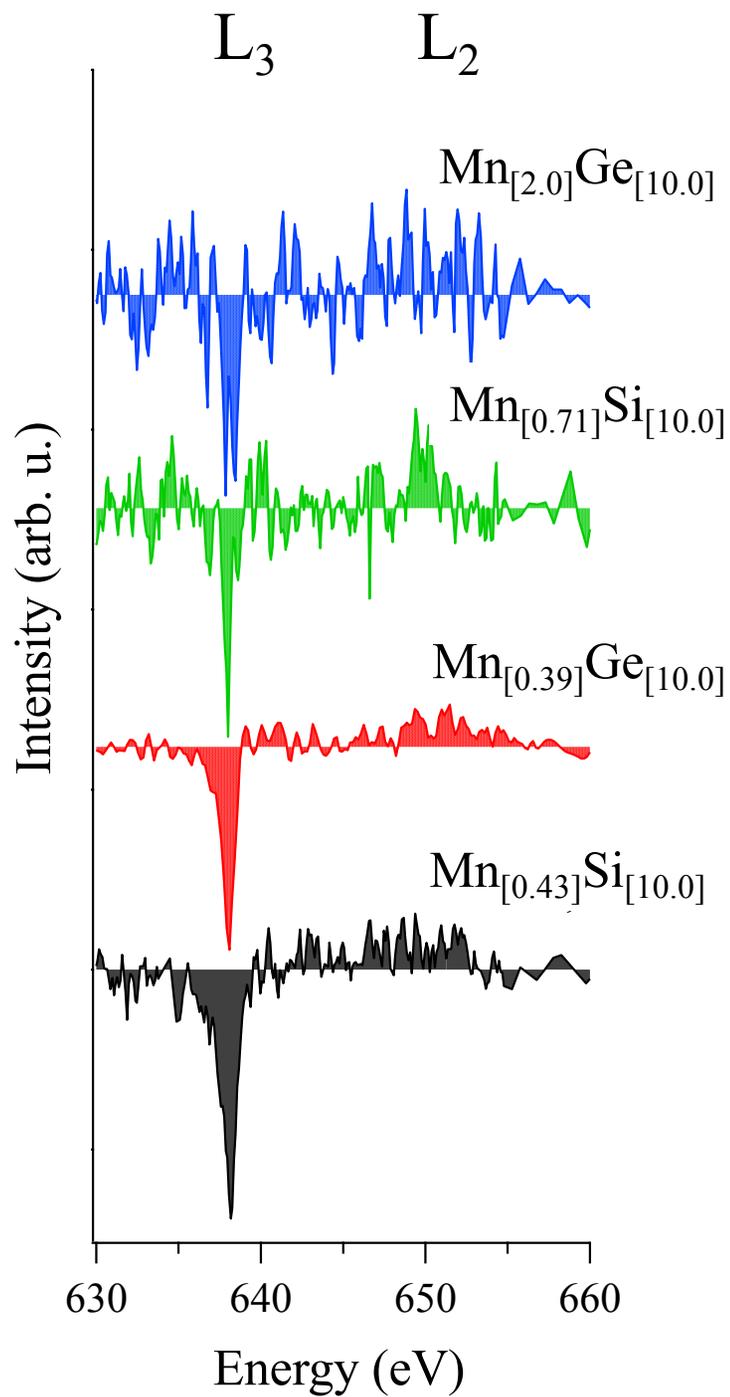
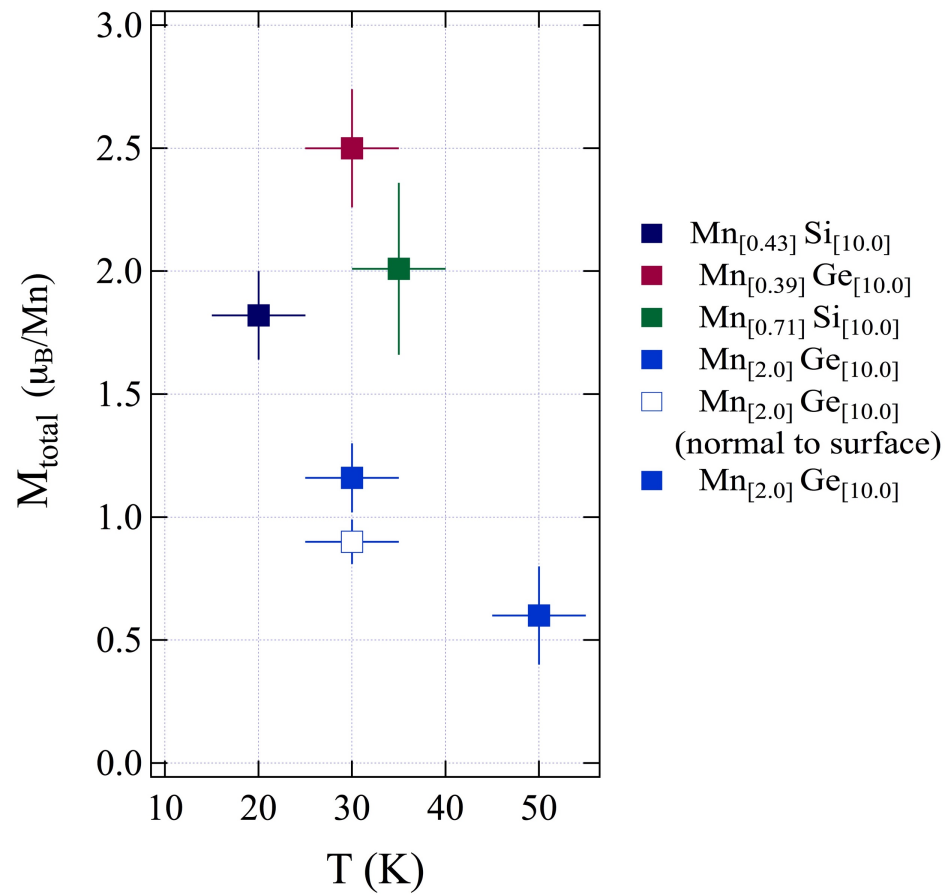